\documentclass[%
 amsmath,amssymb,
 aps,
 prl,
 twocolumn,
]{revtex4-2}
\usepackage[english]{babel}
\usepackage{bibunits}
\usepackage[utf8]{inputenc}
\usepackage[colorinlistoftodos, color=green!40, prependcaption]{todonotes}
\usepackage{amsthm}
\usepackage{mathtools}
\usepackage{physics}
\usepackage{xcolor}
\usepackage{graphicx}
\usepackage{adjustbox}
\usepackage{placeins}
\usepackage[T1]{fontenc}
\usepackage{lipsum}
\usepackage{csquotes}
\usepackage{xcolor}

\usepackage[normalem]{ulem} 
\usepackage[pdftex, pdftitle={Article}, pdfauthor={Author}]{hyperref} 
\begin{document}
\preprint{APS/123-QED}
\title{Engineering Purely Nonlinear Coupling with the Quarton}

\author{Yufeng Ye$^{1,2}$}
\author{Kaidong Peng$^{1,2}$}
\author{Mahdi Naghiloo$^{2}$}
\author{Gregory Cunningham$^{2,3}$}
\author{Kevin P. O'Brien$^{1,2}$}
    
\email[Correspondence email address: ]{kpobrien@mit.edu}%

\affiliation{$^1$Department of Electrical Engineering and Computer Science, Massachusetts Institute of Technology, Cambridge, MA 02139, USA}
\affiliation{$^2$Research Laboratory of Electronics, Massachusetts Institute of Technology, Cambridge, MA 02139, USA}
\affiliation{$^3$Harvard John A. Paulson School of Engineering and Applied Sciences, Harvard University, Cambridge, MA 02138, USA}

\date{\today} 

\begin{abstract}
Strong nonlinear coupling of superconducting qubits and/or photons is a critical building block for quantum information processing. 
Due to the perturbative nature of the Josephson nonlinearity, linear coupling is often used in the dispersive regime to approximate nonlinear coupling. However, this dispersive coupling is weak and the underlying linear coupling mixes the local modes which, for example, distributes unwanted self-Kerr to photon modes. 
Here, we use the quarton to yield purely nonlinear coupling between two linearly decoupled transmon qubits.
The quarton’s zero $\phi^2$ potential enables a giant gigahertz-level cross-Kerr which is an order of magnitude stronger compared to existing schemes, 
and the quarton's positive $\phi^4$ potential can cancel the negative self-Kerr of qubits to linearize them into resonators. This giant cross-Kerr between bare modes of qubit-qubit, qubit-photon, and even photon-photon is ideal for applications such as single microwave photon detection and implementation of bosonic codes. 
\end{abstract}

\maketitle

    \textit{Introduction.}--Circuit quantum electrodynamics (cQED) with microwave superconducting circuits is at the forefront of quantum information processing \cite{yaleReview, supremacy}. In this platform, the nonlinearity of the Josephson junction (JJ) fulfills two fundamental purposes: 
    (1) to provide a self-Kerr nonlinearity which turns otherwise linear resonators into nonlinear artificial atoms that serve as qubits \cite{alexandre, BlaisReview2020}, and 
    (2) to provide a nonlinear coupling between modes, which enables non-classical interactions such as qubit-qubit (matter-matter) \cite{SinglePhotonTransistor, JRMannealer} and photon-qubit (light-matter) \cite{EngineersGuide, BlaisReview2020} entanglement, squeezing \cite{ArneSqueezing}, and amplification \cite{jtwpa}. However, like most nonlinear phenomena in nature \cite{nonlinear-limit}, the JJ's nonlinear response is perturbative (weak) relative to its linear response in the single-photon regime.
    It is therefore difficult to achieve purely nonlinear coupling \cite{SinglePhotonTransistor, JRMannealer, PRX_crossKerr_readout} without an accompanying and often undesirable linear coupling.
    As such, \textit{linear} interactions like the capacitive coupling between a transmon \cite{transmon} and a resonator, $g(\hat{a} + \hat{a}^\dagger)(\hat{b} + \hat{b}^\dagger)$, are used in the dispersive limit to \textit{approximate} cross-Kerr type nonlinear coupling of the form $g_{az} \hat{a}^\dagger \hat{a} \hat{\sigma}_z $ \cite{PRX_crossKerr_readout, FanPRL}. 
    The perturbative nature of this dispersive cross-Kerr is limiting in terms of both strength and performance (due to nonidealities) in areas like quantum non-demolition qubit readout \cite{didierReadout, RWA-readout, PRX_crossKerr_readout} and microwave photon detection \cite{FanPRL, PRL2014-QND-PhotonDetection}. 
    Furthermore, linear coupling hybridizes otherwise localized excitations into normal modes distributed among the coupled circuits \cite{BlackBox_Quantization, SipePRL2019}, which impedes local control and tuning of subsystems in devices like amplifiers and detectors \cite{Sipe_OSA_2020, Arne}. In the case of resonator-qubit coupling, the distributed mode of the bare resonator inherits some self-Kerr nonlinearity from the qubit \cite{Nature_SinglePhoton_Kerr, BlaisReview2020}, which is detrimental to the performance of qubit-readout \cite{cancelKerr-readout}, bosonic qubits \cite{100-Photon-CatStates, Shruti-KerrCat}, and their gates \cite{ArnePRX, CatGate_selfKerr, TwoCatGate, selfKerr-teleportation}.

    In this work, we propose purely nonlinear couplers based on a superconducting qubit dubbed the ``quarton'' \cite{quarton}. The quarton was recently demonstrated as a flux qubit with high anharmonicity and long coherence times \cite{quarton}; devices with similar quartic potentials were proposed previously as a highly anharmonic phase qubit with efficient readout \cite{zorin} and demonstrated as part of a superinductor \cite{bell}. 
    The quarton has two key properties relevant to this paper: first, it is itself a purely nonlinear element with \textit{no linear inductance} ($\phi^2$ potential);
    second, it has a \textit{positive} $\phi^4$ nonlinear potential, in contrast with the \textit{negative} $\phi^4$ of the JJ and most JJ-based elements ($\phi$ being superconducting phase). 
    The zero $\phi^2$ of our proposed quarton coupler can enable a giant, purely nonlinear coupling that is an order of magnitude higher compared to existing purely nonlinear couplers like the C-shunt SQUID and Josephson ring modulator (JRM) \cite{SinglePhotonTransistor, photonsolid, npjQI2018, CSQUID-photons, JRMannealer, 3lvJRM, PRX_crossKerr_readout, Transmon_Coupler} which have coupling strengths limited to the self-Kerr of modes.
    Furthermore, the quarton's positive $\phi^4$ nonlinearity can be used to cancel the negative self-Kerr arising from JJs, thereby enabling coupling to linear photonic modes without the need for additional self-Kerr cancellation methods \cite{Fluxonium-KerrCancel, selfKerr-teleportation, Fluxonium-Transmon-Kerr-Cancel}\footnote{Alternatively, Kerr-free three-wave mixing nonlinear elements like the SNAIL\cite{SNAIL} can be used at the expense of forgoing four-wave interactions like the cross-Kerr.}. 
    We can thus achieve giant, purely nonlinear coupling between any combinations of light and matter modes -- beyond the traditional light-matter and matter-matter coupling -- and reach a new regime of light-light purely nonlinear coupling.
    This quarton-based purely nonlinear coupling of bare modes makes the identification and local control of modes easy \cite{SipePRL2019}, enabling applications such as four-wave mixing with pump and signal separation \cite{Sipe_OSA_2020}. Furthermore, the self-Kerr cancellation with giant cross-Kerr could improve bosonic qubit control \cite{selfKerr-teleportation,Fluxonium-KerrCancel,Fluxonium-Transmon-Kerr-Cancel}, rapid read-out and gate schemes \cite{PRX_crossKerr_readout,optical-zero-SPM}, broadband single microwave photon detection \cite{Arne}, quantum annealing architectures \cite{JRMannealer,npj2017_annealing}, and all-microwave control \cite{SinglePhotonTransistor,microwave-flipflop}.
    
\begin{figure}
\includegraphics[width=0.48\textwidth]{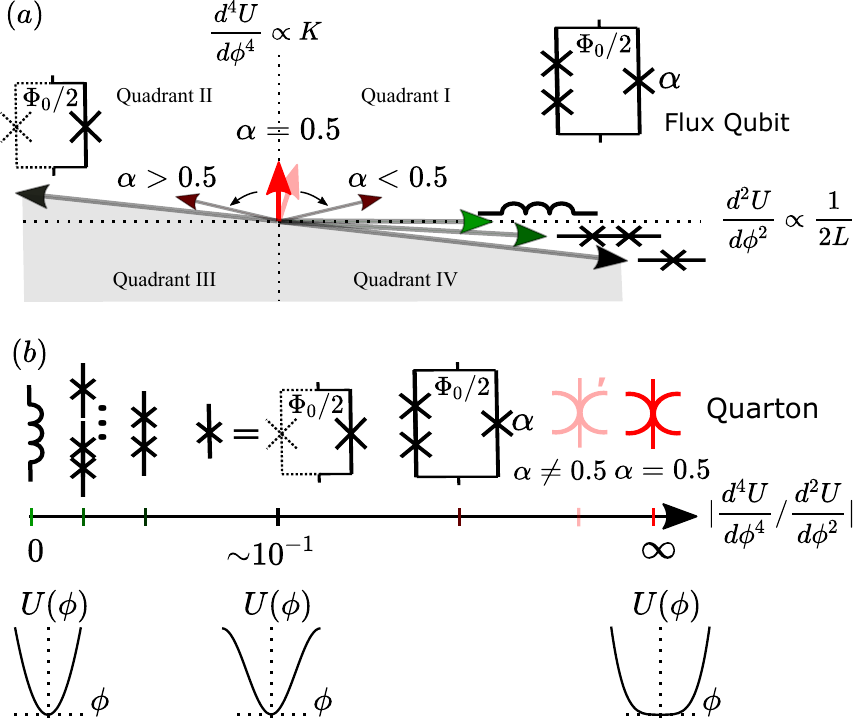}
\caption{\label{fig:fig1} The quarton as a purely nonlinear element. (a) Schematic plot of the nonlinear ($K$) vs linear ($\frac{1}{2L}$) landscape of inductive superconducting elements with centrosymmetric potentials ($U$). The flux qubit has a negative and a positive inductance branch with strength dependent on ratio $\alpha$. The quarton (red) is a special flux qubit with $\alpha = 0.5$ that has no linear potential. The grey region is energetically unstable. (b) Schematic line scale of the relative nonlinearity of the elements in (a). The quarton (red spider symbol) is at infinity. The respective potentials $U(\phi)$'s are plotted below. The tilted quarton (light red spider symbol with apostrophe) is a quarton with small linear inductive potential.}
\end{figure}

\textit{The quarton.}--We start by categorizing superconducting circuit elements by their nonlinearities, which are usually derived from the cosine potential of the JJ \cite{tinkham}: 
\begin{equation}
 U_{JJ}(\phi) = -E_J \cos{\phi} \approx \frac{E_J}{2} \phi^2 - \frac{E_J}{24} \phi^4 \approx \frac{\phi_0^2}{2 L_J} \phi^2 - K \phi^4
 \label{eq:Cos},
\end{equation}
where $E_J$ is the Josephson energy. Physically, JJs with superconducting phase $\phi$ exhibit both a positive linear inductance $L_J$ given by the quadratic $\phi^2$ component of the potential and a negative nonlinear inductance given by the quartic $\phi^4$ (and higher order) component of the potential. For the remainder of the paper, we assume $\phi \ll 1$ and keep up to the quartic $\phi^4$ term which is characterized by the nonlinear Kerr coefficient $K$. 

We repeat this for a wide range of inductive superconducting elements, which leads to a schematic plot of their nonlinear ($\frac{d^4U}{d\phi^4}$) vs linear ($\frac{d^2U}{d\phi^2}$) energy coefficient  in Fig.~\ref{fig:fig1}a
\footnote{Note that we include only symmetric potentials and neglect higher-order nonlinearities for simplicity.}. 
For ease of comparison, the slope $|\frac{d^4U}{d\phi^4} / \frac{d^2U}{d\phi^2}|$ is plotted in Fig.~\ref{fig:fig1}b with the potential diagrams $U(\phi)$ illustrated at the bottom.
Following Eq.~(\ref{eq:Cos}), we place the JJ as a vector in quadrant IV of Fig.~\ref{fig:fig1}a with a length proportional to $E_J$. However, the slope or direction of the JJ vector which characterizes its relative nonlinearity is invariant with $E_J$. We can thus think of the linear-nonlinear plane (Fig.~\ref{fig:fig1}a) as a two-dimensional vector space, with different circuit elements as vectors having $E_J$ dependent length but unique directionality (i.e. they have characteristic unit vectors).
Note that Fig.~\ref{fig:fig1} presents only the potential energy of inductors within which the kinetic energy of capacitors can be added (e.g. effectively single JJ qubits like the transmon would have potential energy represented by the JJ vector)\cite{Koch}. 

We consider three techniques that change the relative nonlinearity: (i) add more JJs in series to decrease the relative nonlinearity \cite{DevoretSuperinductor, JAMPA} (ii) thread half a flux quantum ($\Phi_0/2$) of external magnetic flux through a loop of elements, (iii) connect inductive elements in parallel to add their vectors on Fig.~\ref{fig:fig1}a. 
For (i) with $n$ identical JJs in series (all with $E_J \gg E_C$, $E_C$ being capacitive energy \cite{tinkham}), the phase $\phi$ across the chain of JJs is divided evenly across each JJ ($\phi \rightarrow \phi / n$) \cite{DevoretSuperinductor}. By Eq.~(\ref{eq:Cos}), this implies that: $\frac{1}{L} \rightarrow \frac{n}{L}\frac{1}{n^2}, K \rightarrow n K \frac{1}{n^4}$, so more JJs in series lowers $|\frac{d^4U}{d\phi^4} / \frac{d^2U}{d\phi^2}|$. In the limit $n \rightarrow \infty$, we get a superinductor \cite{DevoretSuperinductor}, which is purely linear. 
We can also (ii) add a $\Phi_0 / 2$ flux bias: 
For a multi-branch element like a SQUID, the external flux acts on one JJ branch \cite{koch_flux} and shifts its cosine potential $U_{JJ}(\phi) \rightarrow -U_{JJ}(\phi)$ via $\phi \rightarrow \phi + \pi$. This flips one branch \cite{koch_flux} JJ vector to the quadrant II of Fig.~\ref{fig:fig1}a. 
We can further use (iii) to add vectors to produce devices such as flux qubits that live in the space between the flux-biased and unbiased SQUIDs / JJs. This is valid because flux qubits in general have two parallel branches with the same $\phi$, so the overall potential $U(\phi)$ is a sum of the two branch $U$'s. 

The top right corner of Fig.~\ref{fig:fig1}a shows a conventional flux qubit \cite{quarton} with two identical JJs with $E_J$ in series in one branch, and a smaller area JJ with $\alpha E_J$ in the other branch.
Without loss of generality \cite{koch_flux}, we choose the gauge such that the $\alpha E_J$ branch is flux-biased (quadrant II of Fig.~\ref{fig:fig1}a) and the series JJ branch is unchanged (quadrant IV of Fig.~\ref{fig:fig1}a). 
Because the two branches have different $|\frac{d^4U}{d\phi^4} / \frac{d^2U}{d\phi^2}|$ (Fig.~\ref{fig:fig1}b), the resulting flux qubit vector from the addition of the two branch vectors can have different directions depending on $\alpha$. For instance, the persistent current flux qubit with double well potential \cite{persistent-current} has $\alpha > 0.5$ and lives in quadrant II; whereas the C-shunt flux qubit with single well potential \cite{FluxQubit-revisited} has $\alpha < 0.5$ and lives in quadrant I. Flux qubits with more ($n \geq 2$) series JJs follow the same principle, with potential:
\begin{equation}
    U(\phi) = -n E_J \cos(\frac{\phi}{n}) - \alpha E_J \cos(\phi - \pi)
    \label{eq:flux_qubit}.
\end{equation}
The quarton is the special flux qubit with $\alpha = 0.5$ ($= \frac{1}{n}$ in general), for which the negative inductance from the quadrant II vector exactly cancels the positive inductance from the quadrant IV vector while the stronger positive $K$ of the quadrant II vector survives the addition. The quarton is named after the resulting leading order positive quartic $\phi^4$ potential and zero $\phi^2$ potential. Physically, the linearized current flow in the two branches destructively interfere.
From its position on Fig.~\ref{fig:fig1}a, it is clear that the quarton forms a natural basis in combination with the inductor. In Fig.~\ref{fig:fig1}b, the quarton (spider symbol \cite{spider}) defines the infinity end of the $|\frac{d^4U}{d\phi^4} / \frac{d^2U}{d\phi^2}|$ scale opposite to the linear inductor which defines the zero point. Graphically, the potential diagrams below Fig.~\ref{fig:fig1}b show a completely anharmonic quartic potential for the quarton. See supplemental material for a discussion on a nonlinear optics analogy and the energetically unstable grey region in Fig.~\ref{fig:fig1}a.  

In practice, JJ based inductive elements like the quarton have accompanying junction capacitances which can cause linear capacitive coupling. To mitigate this, we use a slightly linear quarton, dubbed the ``tilted quarton'' for its position in Fig.~\ref{fig:fig1}a (light red). The tilted quarton has some small linear inductive potential which can cancel the coupling effects of an equally small amount of accompanying linear capactance \cite{SinglePhotonTransistor, npjQI2018}. To distinguish it from the quarton, we give the tilted quarton a lightly shaded spider symbol with an extra apostrophe (Fig.~\ref{fig:fig1}b).

\begin{figure}
\includegraphics[width=0.45\textwidth]{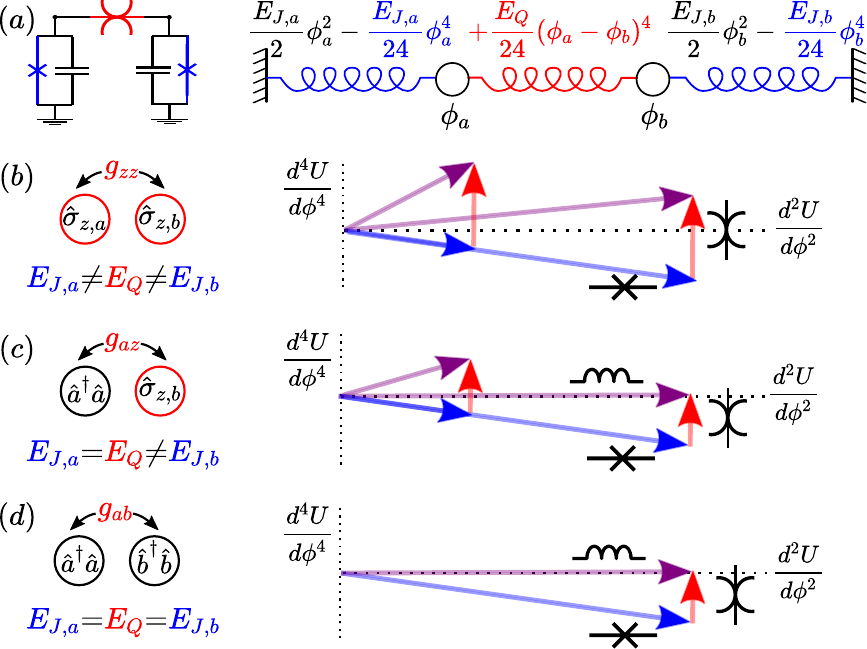} \caption{\label{fig:coupling} Quarton-mediated purely nonlinear light and/or matter interactions. (a) Canonical circuit of quarton (red) nonlinearly coupling two transmons (blue) indexed $a,b$, and its spring-mass analog with Josephson energies as nonlinear spring coefficients $E_{J,a}, E_Q, E_{J,b}$.
Strong nonlinear coupling between (b) Matter-matter modes when ($E_{J,a} \hspace{-3pt}\neq\hspace{-3pt} E_Q \hspace{-3pt}\neq\hspace{-3pt} E_{J,b}$) quarton induces positive qubit nonlinearities, (c) Light-matter modes when ($E_{J,a}{=}E_Q\hspace{-3pt}\neq\hspace{-3pt}E_{J,b}$) quarton cancels qubit $a$'s self-Kerr, (d) Light-light modes when ($E_{J,a}{=}E_Q{=}E_{J,b}$) quarton cancels both qubits' self-Kerr. Conventions: photon annihilation operators - $\hat{a},\hat{b}$, qubit operators - $\hat{\sigma}_{z,ab}$.}
\end{figure}

\textit{Purely nonlinear light and/or matter coupling.}--Consider the canonical circuit of two qubits (labelled $a$ and $b$) coupled via a quarton, shown in Fig.~\ref{fig:coupling}a. 
We can construct an exact spring-mass analogy for the system wherein $\phi, E_J$ are analogs of position and spring constant, respectively. ($E_Q \equiv E_J \frac{n^2-1}{n^3}$ is the effective Josephson energy of the quarton, see supplement for details.)
Note that because there is no linear coupling potential of the form $\frac{E_Q}{2}(\phi_a - \phi_b)^2$ in the red quarton spring, the quarton naturally facilitates purely nonlinear coupling without linear coupling. 

Remarkably, by simply adjusting the relative magnitudes of the qubit spring constants ($E_{J,a}, E_{J,b}$) to the coupling quarton spring constant ($E_Q$), we can access nonlinear coupling between all three combinations of light and matter modes. 
As shown in Fig.~\ref{fig:coupling}b-d, these combinations are longitudinal qubit-qubit coupling ($g_{zz}$), AC Stark shift-like  qubit-photon coupling ($g_{az}$), and cross-Kerr photon-photon coupling ($g_{ab}$), respectively. In particular, Fig.~\ref{fig:coupling}d represents (to our best knowledge) the first system that exhibits \textit{cross-Kerr without self-Kerr} or \textit{photon-photon purely nonlinear coupling}. This is in stark contrast with previous purely nonlinear coupling schemes \cite{SinglePhotonTransistor, JRMannealer, SipePRL2019} that leave modes with non-zero self-Kerr \cite{CSQUID-photons}.  

The quarton's purely nonlinear coupling potential, $\frac{E_Q}{24}(\hat{\phi}_a - \hat{\phi}_b)^4$ can be expanded into:
\begin{equation}
    \frac{E_Q}{24} (\hat{\phi}_a - \hat{\phi}_b)^4 = \frac{E_Q}{24} [\hat{\phi}_a^4 + \hat{\phi}_b^4 + 6 \hat{\phi}_a^2 \hat{\phi}_b^2 - 4(\hat{\phi}_a^3\hat{\phi}_b + \hat{\phi}_a\hat{\phi}_b^3)]
    \label{eq:Qcoupling}.
\end{equation}
 After quantizing $\hat{\phi}_a = \phi_{ZPF,a} (\hat{a} + \hat{a}^\dagger)$, $\hat{\phi}_b = \phi_{ZPF,b} (\hat{b} + \hat{b}^\dagger)$, the $\hat{\phi}_a^2 \hat{\phi}_b^2$ term leads to the important cross-Kerr ($\hat{a}^\dagger \hat{a} \hat{b}^\dagger \hat{b}$) type nonlinear coupling and the $\hat{\phi}_{a,b}^3\hat{\phi}_{b,a}$ terms induce other four wave mixing nonlinear effects including correlated photon hopping \cite{npjQI2018}, third harmonic generation \cite{Boyd}, parametric amplification and squeezing \cite{jtwpa}. Importantly, the positive, non-coupling terms $+\frac{E_Q}{24} \hat{\phi}_{a,b}^4$ can be grouped with the qubits' negative nonlinear potentials $ - \frac{E_{J,ab}}{24} \hat{\phi}_{a,b}^4$ to produce effective qubit nonlinear potentials of $\frac{(E_Q - E_{J,ab})}{24}\hat{\phi}_{a,b}^4$. This can be intuitively represented on the linear-nonlinear diagram for each case. As shown in Fig.~\ref{fig:coupling}b-d, when $E_{Ja,b} \hspace{-3pt}\neq\hspace{-3pt} E_Q$, the vector sum (purple) of the quarton vector (red) and the JJ vector (blue) is non-zero in the nonlinear axis; this represents residual resonator self-Kerr in a qubit mode. In contrast, when $E_{Ja,b}{=}E_Q$, the quarton's induced positive self-Kerr cancels the JJ's intrinsic negative self-Kerr and the resulting sum is zero in the vertical nonlinear axis; this represents a linear resonator photon mode. We emphasize that \textit{in-situ} tuning of $E_{J,a}, E_{J,b}$ is possible with flux-tunable transmons \cite{Koch}. Therefore, results here enable flexible superconducting architectures with modes that can be tuned in-situ to behave either as qubits or resonators.   

\begin{figure}
\includegraphics[width=0.45\textwidth]{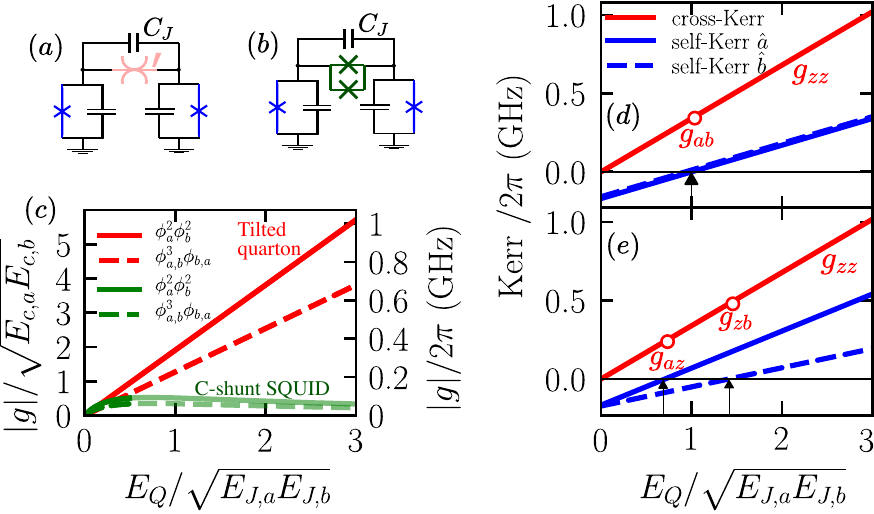}
\caption{\label{fig:3} Purely nonlinear coupling (both $\phi_a^2 \phi_b^2$ and $\phi_{a,b}^3 \phi_{b,a}$)  mediated by (a) tilted quarton (red) with $C_J = 5 $ fF, versus (b) C-shunt SQUID \cite{SinglePhotonTransistor} (green). (c) Nonlinear coupling coefficient $g$ (for $\hat{a}^\dagger \hat{a} \hat{b}^\dagger \hat{b}$ in $\phi_a^2 \phi_b^2$ and similar for $\phi_{a,b}^3 \phi_{b,a}$) scales linearly with $E_Q$ for tilted quarton, allowing for order of magnitude improvement over C-shunt SQUID at large $E_Q$. C-shunt SQUID's linear coupling cancellation relies on RWA which is invalid for large $E_Q$ (light green). For tilted quarton: (d) Simultaneous self-Kerr ($(\hat{a}^\dagger)^2 \hat{a}^2, (\hat{b}^\dagger)^2 \hat{b}^2$) cancellation is possible with $E_{J,a}{=}E_{J,b}$, which is also used in (c). (e) Same qubits flux-tuned to $E_{J,a} \hspace{-3pt}\neq\hspace{-3pt} E_{J,b}$ leads to self-Kerr cancellation of only one mode at a time.}
\end{figure}

We contrast potential realizations of the canonical circuit in Fig.~\ref{fig:coupling}a with two state-of-art nonlinear couplers: the C-shunt SQUID \cite{SinglePhotonTransistor, photonsolid, npjQI2018, CSQUID-photons} which cancels inductive and capacitive linear coupling within the rotating wave approximation (RWA), and the Josephson ring modulator (JRM) \cite{JRMannealer, 3lvJRM} which cancels all linear coupling as well as asymmetric nonlinear coupling ($\phi_{a,b}^3 \phi_{b,a}$) terms by symmetry. The two qubits $a, b$ to be coupled are typical transmons and properties are calculated both analytically and numerically using QuCAT \cite{qucat1}. See supplemental material for related derivations and calculations. 

Analogous to the C-shunt SQUID, we use the tilted quarton (Fig.~\ref{fig:3}a) to cancel (up to RWA) the linear coupling due to intrinsic junction capacitances $C_J$. Unlike the C-shunt SQUID (Fig.~\ref{fig:3}b) which needs a large, variable shunt capacitance $C_J$ to cancel the SQUID inductance, the tilted quarton has intentionally added inductance to the quarton to cancel a small, fixed $C_J$. Henceforth, we use $E_Q$ to denote both the quarton's and the corresponding C-shunt SQUID/JRM's Josephson energy. 
In Fig.~\ref{fig:3}c, we show that for large $E_Q$, quarton-enabled nonlinear coupling strength $g$ (for $\hat{a}^\dagger \hat{a} \hat{b}^\dagger \hat{b}$ in $\phi_a^2 \phi_b^2$) can be an order of magnitude (1 GHz vs 100 MHz) higher than the C-shunt SQUID which limit $g$ to much less than the anharmonicities $E_{c,a}, E_{c,b}$ of the transmons.
This is because all existing couplers have linear inductive potentials which increase $E_{J,ab}$ of the qubits to an effective $E_Q + E_{J,ab}$; or in the spring-mass analogy (Fig.~\ref{fig:coupling}a) the $a,b$ masses oscillate in a stiffer $\frac{E_{J,ab} + E_Q}{2}\phi_{a,b}^2$ spring potential if $\frac{E_Q}{2}(\phi_a-\phi_b)^2$ exists. The stiffer spring reduces oscillation amplitude, or the zero point fluctuation $\phi^4_{ZPF,ab} = \frac{2E_{C,ab}}{ E_{J,ab}} \rightarrow \frac{2E_{C,ab}}{ E_Q + E_{J,ab}}$ quantum mechanically, which directly reduces the coupling \cite{SinglePhotonTransistor, JRMannealer}:
\begin{equation}g \text{(non-quarton)} \propto  \frac{\sqrt{E_{c,a}E_{c,b}} E_{Q}}{\sqrt{E_{Q} + E_{J,a}} \sqrt{E_Q + E_{J,b}}}
\label{eq:gzz_L},
\end{equation}
which has $\lim\limits_{E_Q \gg E_{J,ab}}  g \leq \sqrt{E_{c,a}E_{c,b}}$.
Using the quarton instead, we can avoid the detrimental linear inductance induced $E_{J,ab} \rightarrow E_Q + E_{J,ab}$, and achieve:
\begin{equation}g\text{(quarton)} \propto \sqrt{E_{c,a}E_{c,b}} \frac{E_{Q}}{\sqrt{E_{J,a}} \sqrt{E_{J,b}}}
\label{eq:gzz_noL},
\end{equation}
which offers approximately linear scaling $g \propto E_Q$. See supplemental material for a detailed derivation of Eq.~(\ref{eq:gzz_L}-\ref{eq:gzz_noL}) and the limit to Eq.~(\ref{eq:gzz_noL}).
\begin{figure}
\includegraphics[width=0.45\textwidth]{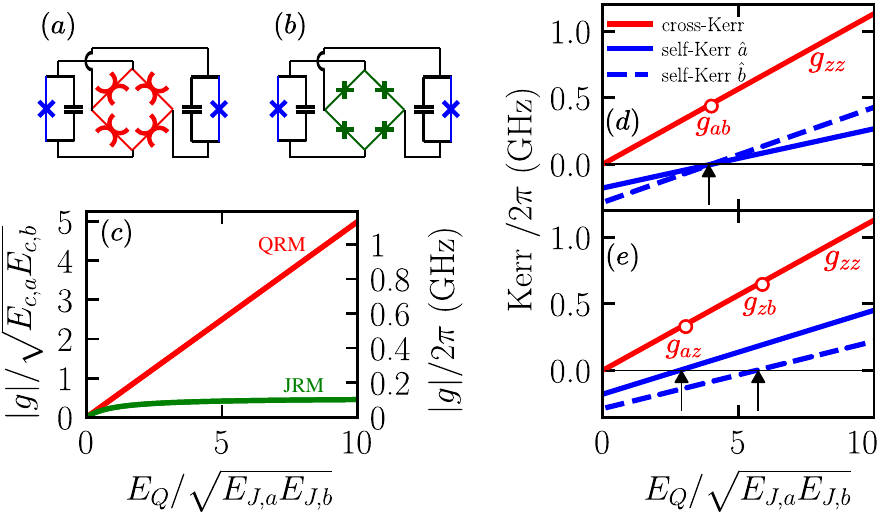}
\caption{\label{fig:4} Purely cross-Kerr ($\phi_a^2 \phi_b^2$) type nonlinear coupling mediated by (a) quarton ring modulator (QRM, red), versus (b) Josephson ring modulator \cite{JRMannealer} (JRM, green). (c) Cross-Kerr strength $g$ (for $\hat{a}^\dagger \hat{a} \hat{b}^\dagger \hat{b}$ in $\phi_a^2 \phi_b^2$) scales linearly with $E_Q$ for QRM, allowing for order of magnitude improvement over JRM at large $E_Q$. For QRM: (d) Simultaneous self-Kerr ($(\hat{a}^\dagger)^2 \hat{a}^2, (\hat{b}^\dagger)^2 \hat{b}^2$) cancellation is possible with $E_{J,a}{=}E_{J,b}$, which is also used in (c). (e) Same qubits flux-tuned to $E_{J,a} \hspace{-3pt}\neq\hspace{-3pt} E_{J,b}$ leads to self-Kerr cancellation of only one mode at a time.}
\end{figure}

In Fig.~\ref{fig:3}d-e, we examine the self-Kerr $(\hat{a}^\dagger)^2 \hat{a}^2, (\hat{b}^\dagger)^2 \hat{b}^2$ (blue) and cross-Kerr $\hat{a}^\dagger \hat{a} \hat{b}^\dagger \hat{b}$ (red) magnitudes for the qubits coupled by tilted quarton and show that all three regimes in Fig.~\ref{fig:coupling}b-d can be reached. For two identical qubits which guarantees $E_{J,a}{=}E_{J,b}$ (Fig.~\ref{fig:3}d), we have that the quarton cancels both qubit self-Kerrs when $E_Q{=}E_{J,a}{=}E_{J,b}$ (black arrow). At that point, there still exists a relatively large $g_{ab}$, enabling strong single photon-photon interactions. With the same qubits flux-tuned to different $E_{J,a} \hspace{-3pt}\neq\hspace{-3pt} E_{J,b}$ (Fig.~\ref{fig:3}e), we can have qubit-photon nonlinear couplings $g_{az}, g_{zb}$ at $E_Q{=}E_{J,a}\hspace{-3pt}\neq\hspace{-3pt}E_{J,b}$ and $E_Q{=}E_{J,b} \hspace{-3pt}\neq\hspace{-3pt} E_{J,a}$ (black arrows), respectively. In general when $E_Q \hspace{-3pt}\neq\hspace{-3pt} E_{J,a} \hspace{-3pt}\neq\hspace{-3pt} E_{J,b}$, we have qubit-qubit $g_{zz}$ nonlinear coupling.

We draw a similar comparison in Fig.~\ref{fig:4} by constructing a quarton ring modulator (QRM) in Fig.~\ref{fig:4}a with the same symmetry as the JRM. The symmetry guarantees that only cross-Kerr type ($\phi_a^2 \phi_b^2$) nonlinear coupling terms exist, and junction capacitances $C_J$ do not cause any linear coupling \footnote{Note that with large shunt $C_J$, the QRM can support a quadruple quarton mode in addition to modes $a,b$. This enables a quarton alternative to the transmon based trimon \cite{3lvJRM}, with cross-Kerr coupled quarton-photon-photon or quarton-transmon-photon or quarton-transmon-transmon modes.}. 
In Fig.~\ref{fig:4}c, we find a similar giant (>1 GHz) $g$ reachable via the QRM, which is an order of magnitude improvement over the JRM. Although the JRM can operate with higher $E_Q$ compared to the C-shunt SQUID \cite{SinglePhotonTransistor, JRMannealer}, its linear inductive potential still limits its $g <= \sqrt{E_{c,a} E_{c,b}}$ by Eq.~(\ref{eq:gzz_L}). Self-Kerr cancellation by the QRM (Fig.~\ref{fig:4}de) can lead to not just purely nonlinear but purely cross-Kerr type coupling between any combination of light and matter modes. We emphasize that their non-dispersive nature enables both tilted quarton and QRM to couple arbitrary frequency (e.g. degenerate) qubits, which could alleviate frequency crowding.

\textit{Conclusion--}
We provided a derivation for the quarton as a purely nonlinear superconducting element, which forms a natural basis with the linear inductor for the linear-nonlinear vector space representation of cQED elements. 
We used these results to show that quarton-based couplers (tilted quarton and QRM) can nonlinearly couple linearly decoupled bare modes, and both facilitate giant (>1 GHz) cross-Kerr interactions and cancel the self-Kerr of matter-like modes, causing them to behave more light-like. This ``quartic regime'' of coupling could be well-suited for applications such as single microwave photon detection and bosonic codes.

This work was funded by the MIT Center for Quantum Engineering via support from the Laboratory for Physical Sciences under contract number H98230-19-C-0292. Y. Ye appreciates financial support from the MIT EECS Jin Au Kong fellowship; G. Cunningham acknowledges support from the Harvard Graduate School of Arts and Sciences Prize Fellowship. The authors thank Arne L. Grimsmo, William D. Oliver for fruitful discussions and insightful comments. 

\bibliography{main}

\end{document}


\preprint{APS/123-QED}
\title{Supplemental Material: Engineering Purely Nonlinear Coupling with the Quarton}

\author{Yufeng Ye$^{1,2}$}
\author{Kaidong Peng$^{1,2}$}
\author{Mahdi Naghiloo$^{2}$}
\author{Gregory Cunningham$^{2,3}$}
\author{Kevin P. O'Brien$^{1,2}$}
    
\email[Correspondence email address: ]{kpobrien@mit.edu}%

\affiliation{$^1$Department of Electrical Engineering and Computer Science, Massachusetts Institute of Technology, Cambridge, MA 02139, USA}
\affiliation{$^2$Research Laboratory of Electronics, Massachusetts Institute of Technology, Cambridge, MA 02139, USA}
\affiliation{$^3$Harvard John A. Paulson School of Engineering and Applied Sciences, Harvard University, Cambridge, MA 02138, USA}

\date{\today} 

\maketitle

\subsection{Analogy with Nonlinear Optics}
We can summarize Fig.~1 of main text in the language of nonlinear optics. Superconducting qubits derive their nonlinearity from the nonlinear potential of the JJ, which can be seen as a nonlinear magnetic element. This is exactly analogous with usual atoms that derive their nonlinearity from the nonlinear potential of the atomic electric field. Therefore, borrowing the classification of nonlinear optical materials by their nonlinear electrical susceptibility $\chi^{(3)}$, we can analogously express superconducting qubit properties in terms of their nonlinear magnetic susceptibility:
\begin{equation}
    \chi_m = \mu_r - 1 = \chi_m^{(1)} + \chi_m^{(3)} H^2
\end{equation}
To make the analogy more exact, we can invoke the duality symmetry for electromagnetic waves \cite{duality}. Briefly, this is the symmetry between electric and magnetic fields in
source-free Maxwell's equations. By Noether's Theorem, there is an associated conserved quantity (helicity) which essentially locks the relative magnitude of $E$ and $H$ fields in vacuum. This can be generalized to materials by applying the constitutive relations via the transformation $E \rightarrow \sqrt{\frac{\epsilon}{\mu}}E$ \cite{duality}. Therefore, a strongly magnetic ($\mu \rightarrow \infty$) material can be seen as an epsilon-near-zero (ENZ) $\epsilon \rightarrow 0$ medium.  

The nonlinear optics language allows us to succinctly capture the power of engineered nonlinear atoms. 
Using $\epsilon \xleftrightarrow{\text{duality}} \frac{1}{\mu}$, we see that the magnetic linear and nonlinear axes of Fig.~1a can be seen as being related to the familiar electric $\chi^{(1)}$ and $\chi^{(3)}$ of nonlinear optics. 
Then, the effect of varying $\alpha$ in Fig.~1a can really be seen as varying nonlinearity $\chi^{(3)}$. Remarkably, we can easily access regions of both positive and negative $\chi^{(3)}$, which is extremely convenient for many applications. For instance, one can pick a nonlinearity of the opposite sign to material dispersion to support solitons in both regions of normal and anomalous disperison. Another example would be quasi-phase matching using alternating regions with opposite signs of nonlinearity \cite{Boyd}. 

For the quarton which has $\mu \rightarrow \infty$, we can see it as having $\epsilon \rightarrow 0$ or equivalently $\chi^{(1)} \rightarrow -1$. This is also valid in practice, as the infinite linear inductance of the quarton represents an electric open -- which is equivalent to a zero capacitance capacitor. In summary, the quarton's trivial linear properties and non-zero $\chi^{(3)}$ makes it a ``plug-and-play'' source of nonlinearity that can be added to materials to edit nonlinearity at will without impacting the linear properties. 

\subsection{No Negative Quarton}

Gauge freedom implies that our convenient choice of spanning tree and closure branch does not suffer any loss of generality \cite{koch_flux}. It immediately follows that our geometric derivation in Fig.~1a is \textit{general} and thus within the assumptions of the derivation, the grey region under the JJ line is not accessible. This means that flux qubits and other elements resulting from flux bias and series JJs cannot be used to produce a more negatively nonlinear superconducting element. In other words, \textit{ there is no negative quarton within our framework}.

At first glance, it is surprising that Fig.~1 is not symmetric about the origin. However, there is a simple and intuitive stability argument for this. Superconducting qubit and resonator systems are exactly analogous to nonlinear spring-mass systems. Therefore, we can lean on the classical intuition that a mass cannot be stable in a concave potential function such as that of a negative quarton. As for quadrant IV of Fig.~1a with negative quartic but positive quadratic potential, by simply examining large $\phi$ behavior and invoking periodicity of the potential function, we can see that these potential functions have deep global minimums at $\phi \neq 0$. Therefore, these systems will also be energetically unstable and tend to absorb or emit a flux quantum (i.e. shift $\phi$) to relax into the true ground state. 

\subsection{Quarton Effective Josephson Energy $E_Q$}
We refer to the effective Josephson energy of the quarton as $E_Q$. It is defined as the equivalent Josephson energy $E_J$ of a normal JJ necessary to reach the same Kerr coefficient ($\frac{d^4 U}{d\phi^4}$) as the quarton. %
For a generalized quarton with $n$ series JJ with $E_J$ in one arm, and $\alpha E_J$ in the other, simply following the derivation in Fig.~1 of main text shows that:
\begin{equation}
    E_Q \equiv E_J \frac{n^2 - 1}{n^3}.
\end{equation}

This definition effectively normalizes all Josephson energies in the circuit and thus allows for an useful direct comparison of JJ's $E_J$ and quarton's $E_Q$ to contrast the Kerr effects in the circuit.

In the QuCAT Modelling section, we use the $E_Q$ to define the JJ and inductor model of the quarton.

\subsection{Choice of $C_J = 5$ fF}
We choose a representative value of 5 fF for the total capacitance between the two transmons in the case of tilted quarton coupling. 

Most of $C_J$ is due to quarton junction capacitance. This is because in order to minimize the phase-slip rate $\propto \exp \left[-\left(8 E_{J} / E_{C}\right)^{1 / 2}\right]$ \cite{tinkham}, we impose $E_J \gg E_C$ for all JJs that make up the quarton. To keep $E_{C}=e^{2} / 2 C$ small, there must be a lower bound on junction capacitance $C$ and by extension $C_J$. 

In the literature, values of flux qubit junction capacitance range from 2.7 \cite{Cj2.7fF} to 6 fF \cite{Cj6fF}. Therefore, we choose $C_J = 5 $ fF as a reasonable value to simulate. 

Note that we neglect the parasitic capacitance to ground for the JJ array that make up the quarton. In practice, this capacitance is extremely small (of order 0.1 fF \cite{Parasitic_C_toGround}) compared to $C_J$ \cite{bell}.

\subsection{Derivation of Nonlinear Coupling $g$}

\subsubsection{State-of-art - coupling with linear $L$}

We will provide a quick summary of the derivations of two relevant works in the literature, one using C-shunt SQUID \cite{SinglePhotonTransistor} and one using Josephson Ring Modulator (JRM) \cite{JRMannealer} for nonlinear coupling without linear coupling.

From Fig.~3a, the Lagrangian of C-shunt SQUID coupled qubits (labelled $a,b$) can be written as follows:
\begin{equation}\begin{aligned}
\mathcal{L} &=\frac{C_a}{2} \dot{\phi}_{a}^{2}+\frac{C_{b}}{2} \dot{\phi}_{b}^{2}+\frac{C_{J}}{2}\left(\dot{\phi}_{a}-\dot{\phi}_{b}\right)^{2} \\
&+E_{J a} \cos \left(\frac{\phi_{a}}{\phi_{0}}\right)+E_{J b} \cos \left(\frac{\phi_{b}}{\phi_{0}}\right)+E_{Q} \cos \left(\frac{\phi_{a}-\phi_{b}}{\phi_{0}}\right)
\end{aligned}
\label{eq:SPT}
\end{equation}

C-shunt SQUID treats the SQUID as just a tunable JJ \cite{SinglePhotonTransistor}, so by expanding the SQUID cosine potential to order $\phi^4$ using Eq.~(1) of main text, we see that:
\begin{equation}
\begin{aligned}
    E_{Q} \cos{(\frac{\phi_a - \phi_b}{\phi_0})} \approx -E_{Q} [&(\frac{\phi_a - \phi_b}{\phi_0})^2 / 2 \\
    & - (\frac{\phi_a + \phi_b}{\phi_0})^4 / 24].
    \label{eq:squid}
\end{aligned}
\end{equation}
in which the linear inductance part:
\begin{equation}
    -\frac{E_{Q}}{2} (\frac{\phi_a - \phi_b}{\phi_0})^2 = -\frac{E_{Q}}{2} [(\frac{\phi_a}{\phi_0})^2 + ( \frac{\phi_b}{\phi_0})^2 - \frac{2\phi_a \phi_b}{\phi_0})].
\end{equation}
The last term provides linear inductive coupling (which will be cancelled by an opposite signed capacitive linear coupling). However, the first two terms are not coupling terms but rather cause a change in effective inductance of the qubits. From Eq.~(\ref{eq:SPT}), expanding the qubit $E_{J,ab}$ cosine terms and collecting coefficients, we find the SQUID $E_{Q}$ changes $E_{J,ab}$ of qubits to an effective $E_Q + E_{J,ab}$. 
The lowering of qubit effective inductance causes a lowering of characteristic impedance and thus the qubit zero point fluctuation $\phi_{ZPF,ab} = (2 E_{c,ab} / E_{J,ab})^{1/4}$ \cite{Girvin} is lowered:
\begin{equation}
    \phi_{ZPF,ab} = [ \frac{ 2E_{c,ab}} { (E_Q + E_{J,ab})}]^{1/4}
\end{equation}

The nonlinear coupling $g$ is derived from the quartic term of Eq.~(\ref{eq:squid}), which when expanded has terms in Eq.~(3) of main text. Using standard quantization $\hat{\phi}_{a} = \phi_{ZPF,a} (\hat{a} + \hat{a}^\dagger)$ and $\hat{\phi}_{b} = \phi_{ZPF,b} (\hat{b} + \hat{b}^\dagger)$, we find that all nonlinear coupling terms (e.g. $\hat{a}^\dagger \hat{a} \hat{b}^\dagger \hat{b}$ for $g_{zz}$) scale with both coupling $E_{Q}$ and the phase zero point fluctuations of the qubits $\phi_{ZPF}$ \cite{Girvin}:
\begin{align}
    g_{zz} & = E_{Q} \phi_{ZPF,a}^2\phi_{ZPF,b}^2 \\
    & \approx 2\sqrt{E_{c,a} E_{c,b}} \frac{E_{Q}}{\sqrt{E_{J a}+E_{Q}} \sqrt{E_{J b}+E_{Q}}},
\end{align}
and we have obtained Eqn.(4) of main text.

The derivation is similar for the JRM coupler (Fig.~4b). For qubit modes labelled $a, b$, the JRM with Josephson energy $E_{Q}$ has coupling potential \cite{JRMannealer}:
\begin{align}& 4  E_{Q} \cos \left(\frac{\phi_{a}}{2}\right) \cos \left(\frac{\phi_{b}}{2}\right)\\
& \approx 4  E_{Q} ( 1 -\frac{1}{2} (\frac{\phi_{a}}{2})^2 + ...)( 1- \frac{1}{2}(\frac{\phi_{b}}{2})^2 + ...)
\end{align}
Similar to C-shunt SQUID, we see that the JRM introduces non-coupling inductive terms $\phi_{a,b}^2$ which modify the inductance of the qubits. The rest follows straightforwardly from the C-shunt SQUID, and we get the same scaling of $g_{zz}$ given by Eqn.(4) in the main text. (Except the JRM has effectively decreased $E_Q \rightarrow \frac{E_Q}{4}$ because of flux division across the two series JJ in the JRM modes \cite{3lvJRM}.) 

In general, we see that any coupling element with linear inductance will lower the qubit inductance. This can bee explained in terms of an equivalent circuit model. When connected to a coupling element with linear inductance, the qubit will see an effective parallel inductor which lowers its own inductance. Therefore, all of these schemes are limited to $g <= E_c$.   

\subsubsection{Quarton coupler - coupling without linear $L$}

When we couple with a quarton which is linearly an electrical open circuit, the qubits do not ``see'' any parallel linear inductance (it is linearly electrically isolated). It follows that the quartic coupling potential of the quarton:
\begin{equation}
    E_Q (\phi_a - \phi_b)^4 \approx E_Q (\phi_a^4 + \phi_b^4 + ...)
\end{equation}
has noncoupling terms $\phi_{a,b}^4$ which only modifies the nonlinearity of the qubits. The quadratic noncoupling terms $\phi_{a,b}^2$ introduced by JRM and C-shunt SQUID are completely absent. As a result, the inductance and thus the zero point fluctuation of the qubits are, to first order, unchanged. Therefore, we get
\begin{align}
    g_{zz} & \approx E_{Q} \phi_{ZPF,a}^2\phi_{ZPF,b}^2 \\
    & \approx 2 \sqrt{E_{C,a} E_{C,b}} \frac{E_{Q}}{\sqrt{E_{J a}} \sqrt{E_{J b}}}
    \label{eq:ideal_Quarton}
\end{align}
corresponding to Eqn.(5) in the main text.

\subsubsection{Fundamental $g_{zz} \gg E_C$ limit}
Deep into the $g_{zz} \gg E_C$ limit, second order effects emerge and the quarton's nonlinear modification of the qubit inductance also start to affect the qubit's zero point fluctuation. 

We have neglected these second order effects thus far because they represent a fundamental limit to high $g_{zz}$. Even the most ideal purely $g_{zz}$ interaction would incur such effects. Thus, they impose a tight bound on the highest possible $g_{zz}$ achievable.  Here, we provide a rough estimate of their influence. 

We can write the quarton's noncoupling corrections $E_Q \phi_{a,b}^4$ as:
\begin{equation}
    E_Q \phi_{a,b}^4 \approx E_Q \phi_{ZPF, ab}^2 \phi_{a,b}^2  
\end{equation}
which is valid when the qubits have low population. For typical transmons \cite{Koch}, we have:
\begin{equation}
    \phi_{ZPF, ab}^2 \approx 0.1
    \label{eq:phi_zpf}
\end{equation}

Therefore, Eq.~(\ref{eq:ideal_Quarton}) would be corrected with:
\begin{equation}g_{zz} \approx 2 \sqrt{E_{c,a} E_{c,b}} \frac{E_{Q}}{\sqrt{E_{J a} + 0.1 E_Q} \sqrt{E_{J b} + 0.1 E_Q}}.
\label{eq:final_gzz}
\end{equation}
which will lead to a weaker nonlinear scaling of $g_{zz}$ with $E_Q$ deep in the regime of $g_{zz} \gg E_{C}$. Ultimately, assuming Eq.~(\ref{eq:phi_zpf}), when $E_Q \rightarrow \infty$  Eq.~(\ref{eq:final_gzz}) bounds the largest possible $g_{zz} <= 20 \sqrt{E_{c,a} E_{c,b}}$ or about 3-5 GHz for typical transmon $E_C$. 

\subsection{Tilted Quarton}
In this section, we first provide a simple derivation for the adverse effect of parasitic linear coupling. These findings motivate the use of the tilted quarton. We end by providing a derivation and sample experimental parameters for a tilted quarton coupler.

The additional $C_J$ causes linear coupling which can be effectively modelled by introducing some additional Jaynes-Cummings (J.C.) -like term \cite{BlaisReview2020} into an otherwise ideal cross-Kerr coupled Hamiltonian:
\begin{align}
    H & = \omega_a \hat{a}^\dagger \hat{a} + \omega_b \hat{b}^\dagger \hat{b} + g_{ab} \hat{a}^\dagger \hat{a} \hat{b}^\dagger \hat{b} \\
    & \xrightarrow[]{J.C.} \omega_a \hat{a}^\dagger \hat{a} + \omega_b \hat{b}^\dagger \hat{b} + g_{ab} \hat{a}^\dagger \hat{a} \hat{b}^\dagger \hat{b} + g_c(\hat{a}^\dagger \hat{b} + \hat{b}^\dagger \hat{a})
    \label{eq:JC}.
\end{align}
Clearly, the linear coupling $g_c$ mixes the two modes $a, b$. Physically, this means that the normal modes of the circuit are now no longer locally confined but are instead distributed. Distributed modes resulting from linear coupling tend to experience weaker nonlinear coupling as a result of a decrease in flux across the nonlinear coupler (quarton). In the extreme limit that $C_J \rightarrow \infty$, the $a,b$ modes are completely hybridized and excitation of the normal mode would cause strong oscillations on both sides of the nonlinear coupler. The symmetrized normal mode, in particular, would not be affected by the presence of the quarton at all.

The linear coupling also compromises self-Kerr cancellation. We can see this by diagonalizing the J.C. part to get eigenmodes $\Tilde{a}, \Tilde{b}$ which will be some superposition of the original $a,b$ modes:
\begin{align}
    a & = \xi \Tilde{a} + \eta \Tilde{b} \\
    b & = \eta \Tilde{a} - \xi \Tilde{b}
\end{align}
And we see that this diagonalization turns the cross-Kerr term into:
\begin{equation}
    g_{ab} \hat{a}^\dagger \hat{a} \hat{b}^\dagger \hat{b} \rightarrow g_{ab} (\xi \Tilde{a} + \eta \Tilde{b})^\dagger (\xi \Tilde{a} + \eta \Tilde{b}) (\eta \Tilde{a} - \xi \Tilde{b})^\dagger (\eta \Tilde{a} - \xi \Tilde{b})
\end{equation}
After expansion, this re-introduces self-Kerr terms $\Tilde{a}^{\dagger2} \Tilde{a}^2$ and $\Tilde{b}^{\dagger2} \Tilde{b}^2$ into what was a purely cross-Kerr interaction. 

In general, localized circuits modes with only nonlocal (nonlinear) interactions are much ``cleaner'' and more intuitive systems to work with for many applications \cite{SipePRL2019}. (However, for device designs that have features such as photonic bandgaps that protect against small linear coupling, e.g. the traveling wave photon detector \cite{Arne}, using a tilted quarton is unnecessary.)

Therefore, as we discussed in the main text, the effect of $C_J$ should be cancelled by using a ``tilted'' quarton with the right amount of linear inductive energy. A detailed derivation of how inductive linear coupling $g_l$ and capactive linear coupling $g_c$ have opposite signs and thus cancel can be found in the literature \cite{SinglePhotonTransistor}. As such, we ``tilt'' the quarton's position in Fig.~1a to reintroduce linear inductance such that the condition analogous to that derived in \cite{SinglePhotonTransistor}:
\begin{equation}
    \frac{E_{Q,tilt}}{\sqrt{E_{J a}+E_{Q, tilt}} \sqrt{E_{J b}+E_{Q,tilt}}}=\frac{C_{J}}{\sqrt{C_{a}+C_{J}} \sqrt{C_{b}+C_{J}}}
    \label{eq:EQ_tilt}
\end{equation}
 is satisfied. (Note that this cancellation assumes the rotating wave approximation (RWA) \cite{SinglePhotonTransistor}.) As a result, the J.C. term of Eq.~(\ref{eq:JC}) is gone and a purely nonlinear coupling Hamiltonian can be constructed. Note here that $E_{Q,tilt}$ is an effective Josephson energy corresponding to the positive linear inductance of the tilted quarton $L_{tilt}$. We can straightforwardly rewrite Eq.~(\ref{eq:EQ_tilt}) in terms of $L_{tilt}$ and $g \equiv \frac{C_{J}}{\sqrt{C_{a}+C_{J}} \sqrt{C_{b}+C_{J}}}$:
 
 \begin{equation}
 \begin{aligned}
L_{t i l t}=& \frac{-\left(L_{J a}+L_{J b}\right)}{2} \\
&+\frac{\sqrt{\left(L_{J a}+L_{J b}\right)^{2}-4 L_{J a} L_{J b}\left(1-\frac{1}{g^{2}}\right)}}{2}
\end{aligned}
\end{equation}
 In the limit that $L_{Ja} = L_{Jb} \equiv L_{Jab}$, we get:
 \begin{equation}
     \frac{L_{tilt}}{L_{Jab}} = \frac{1}{g} - 1
     \label{eq:Ltilt_g}
 \end{equation}
 In QuCAT, this $L_{tilt}$ would emerge from the difference between $L_{Jq,pos}$ and $L_{Jq}$:
 \begin{equation}
     \frac{1}{L_{tilt}} = \frac{1}{|L_{Jq,pos}|} - \frac{1}{|L_{Jq}|}
     \label{eq:Ltilt_QUCAT}
 \end{equation}

For realistic parameters like $C_J = 5 $ fF, $C_a = C_b = 95 $ fF, we expect $g = 0.05$ and, by Eq.~(\ref{eq:Ltilt_g}), $L_{tilt} = 19 L_{Jab}$. 

If $L_{Jab} \approx L_{Jq} = 10 $ nH, a $L_{tilt}$ of 190 nH is needed, which by Eq.~(\ref{eq:Ltilt_QUCAT}) leads to $L_{Jq,pos} = 9.5 $ nH. Compare this with the $L_{Jq,pos} = L_{Jq} = 10$ nH of an untilted quarton (Eq.~(\ref{eq:QuCAT_quarton})), we see that a linear coupling g = 5\% can produce a roughly 5\% variation in quarton branch inductance $L_{Jq,pos}$. Such variations are well within state-of-art fabrication constraints \cite{Ej_precision}. 
\subsection{Proposed couplers' reduction to canonical circuit (Fig.~2a of main text).}

In this section, we provide a quick derivation for how the tilted quarton (Fig.~3a of main text) and the QRM (Fig.~4a of main text) couplers reduce to the canonical circuit in Fig.~2a of main text. 

The tilted quarton, as described in the previous section, cancels linear coupling (up to RWA). So the remaining purely nonlinear coupling from the tilted quarton directly leads to the canonical circuit. 

\begin{figure}
\includegraphics[width=0.5\textwidth]{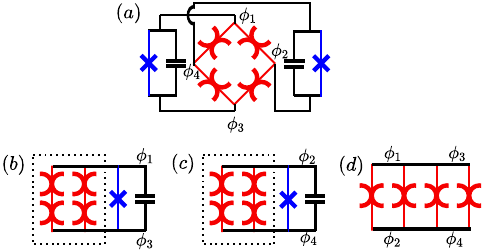}
\centering
\caption{\label{fig:QRM} (a) QRM coupling circuit in the basis $\{\phi_1, \phi_2, \phi_3, \phi_4\}$. (b) Dipole transmon normal mode $\phi_a = \phi_1 - \phi_3$, and (c) dipole transmon normal mode $\phi_b = \phi_2 - \phi_4$. (d) Quadruple quarton normal mode $\phi_c = \phi_1 - \phi_2 + \phi_3 - \phi_4$.}
\end{figure}

The QRM, by symmetry, couples two dipole modes each composed of a transmon in parallel with two branches of two series quartons.
This is shown in Fig.~\ref{fig:QRM}bc, and is exactly analogous to modes shown in Fig.~5 of referenced JRM trimon \cite{3lvJRM}. It is therefore clear that the parallel quarton circuit (dashed box) contributes an effective positive nonlinear potential $\frac{4}{2^4} E_Q/24 \phi_{a,b}^4$ to the transmon, exactly as experienced by the transmon in the canonical circuit except $E_Q \rightarrow E_Q / 4$ due to the weaker nonlinearity of series quarton. 
Note that there is also a quadruple quarton mode in Fig.~\ref{fig:QRM}d, as discussed in main text.  

Not only are the two normal modes of the QRM circuit reducible to the canonical, we can further show that the coupling between them are also similar to the canonical's except the absence of $\phi^3_{a,b}\phi_{b,a}$ coupling terms due to symmetry. 
Consider the potential of the QRM written in the basis $\{\phi_1, \phi_2, \phi_3, \phi_4\}$ as shown in Fig.~\ref{fig:QRM}a:

\begin{equation}
    U_{QRM} = \frac{E_Q}{24}[(\phi_1 - \phi_2)^4 + (\phi_2 - \phi_3)^4 + (\phi_3 - \phi_4)^4 + (\phi_4 - \phi_1)^4]
\end{equation}

We now define new normal mode coordinates:
\begin{equation}
    \begin{aligned}
        \phi_a & = \phi_1 - \phi_3 \\
        \phi_b & = \phi_2 - \phi_4 \\
        \phi_c & = \phi_1 - \phi_2 + \phi_3 - \phi_4
    \end{aligned}
\end{equation}
in which it is easy to verify that:
\begin{equation}
\begin{aligned}
    U_{QRM} = \frac{E_Q}{24 \times 2^4}[&(\phi_a - \phi_b + \phi_c)^4 \\
    & +(\phi_a - \phi_b - \phi_c)^4 \\
    & +(\phi_a + \phi_b - \phi_c)^4 \\
    & +(\phi_a + \phi_b + \phi_c)^4].
    \end{aligned}
\end{equation}
By expanding, we find that all $\phi_{a,b,c}^2$ and $\phi_{a,b}^3 \phi_{b,a}$ terms are cancelled due to symmetry. We are left with:
\begin{equation}
    U_{QRM} = \frac{4}{24 \times 2^4}E_Q [\phi_a^4 + \phi_b^4 + \phi_c^4 + 6 (\phi_a^2\phi_b^2 + \phi_a^2\phi_c^2 + \phi_b^2\phi_c^2)],
\end{equation}
which contains the same $\phi_{a,b}^2 \phi_{b,a}^2$ coupling terms as Eq.~(3) with a weaker $E_Q \rightarrow E_Q / 4$ as before. (Note that if junction capacitances are small, the quadruple quarton mode $\phi_c$ will have high frequency and can thus be ignored \cite{JRMannealer}.) The same symmetry argument can be applied again to show that all linear capacitive couplings between $a,b,c$ modes cancel (see \cite{3lvJRM}). Therefore, the QRM coupling circuit is ultimately equivalent to the canonical circuit with $E_Q \rightarrow E_Q / 4$ and no $\phi_{a,b}^3 \phi_{b,a}$ coupling terms.

\subsection{QuCAT Model}

\begin{figure}
\includegraphics[width=0.35\textwidth]{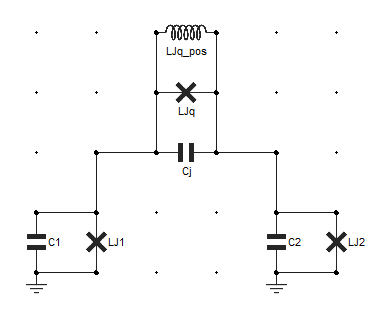}
\centering
\caption{\label{fig:QuCAT} QuCAT circuit schematic for canonical quarton nonlinear coupling of two transmons.}
\end{figure}
In Fig.~\ref{fig:QuCAT}, we show the circuit schematic generated in QuCAT \cite{qucat1} corresponding to the canonical quarton coupling circuit in Fig.~2a of main text. 
On the left and right side, we have the two transmons with capacitance $C_1, C_2$ and JJ linear inductance $L_{J,1}, L_{J,2}$, respectively (index $1,2$ used interchangeably with $a,b$). 
In the center, we have the total capacitance $C_J$ between the qubits (set to 0 for canonical), and the quarton represented by a negative JJ with $L_{Jq}$ in parallel with a positive inductor with $L_{Jq,pos}$.  
Recall from Fig.~1 of main text that the quarton is made up of the interference between a positive and negative inductance branch. Here, for simplicity, we model the positive inductance branch as just an inductor (this corresponds to the number of series JJ $n \rightarrow \infty$ limit). Therefore, by imposing that:
\begin{equation}
L_{Jq} = L_{Jq,pos}
\label{eq:QuCAT_quarton},
\end{equation}
we have a working quarton with $E_Q = |\frac{\phi^2_0}{2L_{Jq}}|$. See the section on tilted quarton for which $|L_{Jq,pos}| < |L_{Jq}|$.

\begin{table*}
\begin{ruledtabular}
\begin{tabular}{cccccc}
&
\textrm{$C_a$}&
\textrm{$C_b$}&
\textrm{$C_J$}&
\textrm{$\omega_{a}$}&
\textrm{$\omega_{b}$}\\
\hline
$g_{ab}$ & 103 fF & 62.5 fF & 5 fF & 7 GHz & 9 GHz \\
$g_{az}, g_{zb}$ & 103 fF & 62.5 fF & 5 fF & 4.95 GHz & 9 GHz \\
\end{tabular}
\caption{\label{tab:Fig3_param}Circuit parameters for Fig.~3 results of main text.}
\end{ruledtabular}
\end{table*}

\begin{table*}
\begin{ruledtabular}
\begin{tabular}{cccccc}
&
\textrm{$C_a$}&
\textrm{$C_b$}&
\textrm{$C_J$}&
\textrm{$\omega_{a}$}&
\textrm{$\omega_{b}$}\\
\hline
$g_{ab}$ & 103 fF & 103 fF & 5 fF & 7 GHz & 7 GHz \\
$g_{az}, g_{zb}$ & 103 fF & 103 fF & 5 fF & 4.95 GHz & 7 GHz \\
\end{tabular}
\caption{\label{tab:Fig4_param}Circuit parameters for Fig.~4 results of main text.}
\end{ruledtabular}
\end{table*}

Using this model for the quarton and transmons, we construct the circuits as defined in Fig.~3ab and Fig.~4ab of main text. In Table \ref{tab:Fig3_param}, we summarize the parameters used for the results in Fig.~3 of main text. In Table \ref{tab:Fig4_param}, we summarize the parameters used for the results in Fig.~4 of main text.

Because QuCAT cannot apply the rotating wave approximation (RWA) necessary for both tilted quarton and C-shunt SQUID models, we calculate results analytically for Fig.~3 of main text. Using the derivation from the previous sections, we have that:

\begin{equation}
\begin{aligned}
    & g_{zz} (\text{tilted quarton}) \\
    & = 2 \sqrt{E_{c,a} E_{c,b}} \frac{E_Q}{\sqrt{ (E_{J,a} + E_{Q,tilt}) (E_{J,b} + E_{Q,tilt})}}
\end{aligned}
\label{eq:g_zz_tQ}
\end{equation}

\begin{equation}
\begin{aligned}
    & g_{zz} (\text{C-shunt SQUID}) \\
    & = 2 \sqrt{E_{c,a} E_{c,b}} \frac{E_Q}{\sqrt{ (E_{J,a} + E_{Q}) (E_{J,b} + E_{Q})}}
\end{aligned}
    \label{eq:g_zz_cS}
\end{equation}

Note that $E_{Q,tilt} = \frac{\phi_0^2}{L_{tilt}}$ is a constant for fixed $C_J = 5$ fF.

Also note that $g_{zz}$ cross-Kerr terms originate from $\phi_a^2 \phi_b^2$. Because we've chosen identical qubits, the other $\phi_{a,b}^3 \phi_{b,a}$ terms are simply a factor of $\frac{4}{6}$ weaker from the coefficient of expansion (Eq.~(3) of main text). 

The self-Kerrs $K_{a,b}$ of tilted quarton coupled transmons are simply the usual transmon self-Kerr \cite{Girvin} modified with effective nonlinear potential $ -E_{J,ab} \rightarrow -E_{J,ab} + E_Q$ and linear potential $E_{J,ab} \rightarrow E_{J,ab} + E_{Q,tilt}$:

\begin{equation}
    K_{a,b} = E_{c,ab} \frac{E_Q - E_{J,ab}}{E_{J,ab} + E_{Q,tilt}}
    \label{eq:K_ab}.
\end{equation}

For Fig.~4 of main text, no RWA is needed so we use the built-in self and cross-Kerr calculation functions in QuCAT to numerically compute all values \footnote{The $L_J$'s of all component JJs (and $L_{Jq,pos}$) in the JRM / QRM must be slightly different to avoid some code errors.}. All numerical results matched exactly with the analytic equations (which is same as Eq.~(\ref{eq:g_zz_tQ}) and Eq.~(\ref{eq:K_ab}) except $E_Q \rightarrow E_Q/4$ and $E_{Q,tilt} = 0$).

Note that because QuCAT does not handle high anharmonicity modes ($\alpha / \omega > 0.06$) well \cite{qucat1}, all our simulation parameters adhered to this constraint by use of low anharmonicity transmons and limited $E_Q$.

\bibliography{sm}